\documentclass[%
 reprint,
superscriptaddress,
 amsmath,amssymb,
 aps
]{revtex4-2}

\usepackage{graphicx}
\usepackage{dcolumn}
\usepackage{bm}
\usepackage{hyperref}
\usepackage[mathlines]{lineno}

\begin{document}


\title{Measurement of the pion form factor with CMD-3 detector and its implication to the hadronic contribution to muon (g-2)}

\author{F.V.~Ignatov}
\email{F.V.Ignatov@inp.nsk.su}
\affiliation{Budker Institute of Nuclear Physics, SB RAS, Novosibirsk, 630090, Russia}
\affiliation{Novosibirsk State University, Novosibirsk, 630090, Russia}

\author{R.R.~Akhmetshin}
\affiliation{Budker Institute of Nuclear Physics, SB RAS, Novosibirsk, 630090, Russia}
\affiliation{Novosibirsk State University, Novosibirsk, 630090, Russia}

\author{A.N.~Amirkhanov}
\affiliation{Budker Institute of Nuclear Physics, SB RAS, Novosibirsk, 630090, Russia}
\affiliation{Novosibirsk State University, Novosibirsk, 630090, Russia}

\author{A.V.~Anisenkov}
\affiliation{Budker Institute of Nuclear Physics, SB RAS, Novosibirsk, 630090, Russia}
\affiliation{Novosibirsk State University, Novosibirsk, 630090, Russia}

\author{V.M.~Aulchenko}
\affiliation{Budker Institute of Nuclear Physics, SB RAS, Novosibirsk, 630090, Russia}
\affiliation{Novosibirsk State University, Novosibirsk, 630090, Russia}

\author{N.S.~Bashtovoy}
\affiliation{Budker Institute of Nuclear Physics, SB RAS, Novosibirsk, 630090, Russia}

\author{D.E.~Berkaev}
\affiliation{Budker Institute of Nuclear Physics, SB RAS, Novosibirsk, 630090, Russia}
\affiliation{Novosibirsk State University, Novosibirsk, 630090, Russia}

\author{A.E.~Bondar}
\affiliation{Budker Institute of Nuclear Physics, SB RAS, Novosibirsk, 630090, Russia}
\affiliation{Novosibirsk State University, Novosibirsk, 630090, Russia}

\author{A.V.~Bragin}
\affiliation{Budker Institute of Nuclear Physics, SB RAS, Novosibirsk, 630090, Russia}

\author{\fbox{S.I.~Eidelman}}
\affiliation{Budker Institute of Nuclear Physics, SB RAS, Novosibirsk, 630090, Russia}
\affiliation{Novosibirsk State University, Novosibirsk, 630090, Russia}

\author{D.A.~Epifanov}
\affiliation{Budker Institute of Nuclear Physics, SB RAS, Novosibirsk, 630090, Russia}
\affiliation{Novosibirsk State University, Novosibirsk, 630090, Russia}

\author{L.B.~Epshteyn}
\affiliation{Budker Institute of Nuclear Physics, SB RAS, Novosibirsk, 630090, Russia}
\affiliation{Novosibirsk State University, Novosibirsk, 630090, Russia}
\affiliation{Novosibirsk State Technical University, Novosibirsk, 630092, Russia}

\author{A.L.~Erofeev}
\affiliation{Budker Institute of Nuclear Physics, SB RAS, Novosibirsk, 630090, Russia}
\affiliation{Novosibirsk State University, Novosibirsk, 630090, Russia}

\author{G.V.~Fedotovich}
\affiliation{Budker Institute of Nuclear Physics, SB RAS, Novosibirsk, 630090, Russia}
\affiliation{Novosibirsk State University, Novosibirsk, 630090, Russia}

\author{A.O.~Gorkovenko}
\affiliation{Budker Institute of Nuclear Physics, SB RAS, Novosibirsk, 630090, Russia}
\affiliation{Novosibirsk State Technical University, Novosibirsk, 630092, Russia}

\author{F.J. Grancagnolo}
\affiliation{%
 Instituto Nazionale di Fisica Nucleare, Sezione di Lecce, Lecce, Italy
}%

\author{A.A.~Grebenuk}
\affiliation{Budker Institute of Nuclear Physics, SB RAS, Novosibirsk, 630090, Russia}
\affiliation{Novosibirsk State University, Novosibirsk, 630090, Russia}

\author{S.S.~Gribanov}
\affiliation{Budker Institute of Nuclear Physics, SB RAS, Novosibirsk, 630090, Russia}
\affiliation{Novosibirsk State University, Novosibirsk, 630090, Russia}

\author{D.N.~Grigoriev}
\affiliation{Budker Institute of Nuclear Physics, SB RAS, Novosibirsk, 630090, Russia}
\affiliation{Novosibirsk State University, Novosibirsk, 630090, Russia}
\affiliation{Novosibirsk State Technical University, Novosibirsk, 630092, Russia}

\author{V.L.~Ivanov}
\affiliation{Budker Institute of Nuclear Physics, SB RAS, Novosibirsk, 630090, Russia}
\affiliation{Novosibirsk State University, Novosibirsk, 630090, Russia}

\author{S.V.~Karpov}
\affiliation{Budker Institute of Nuclear Physics, SB RAS, Novosibirsk, 630090, Russia}

\author{A.S.~Kasaev}
\affiliation{Budker Institute of Nuclear Physics, SB RAS, Novosibirsk, 630090, Russia}

\author{V.F.~Kazanin}
\affiliation{Budker Institute of Nuclear Physics, SB RAS, Novosibirsk, 630090, Russia}
\affiliation{Novosibirsk State University, Novosibirsk, 630090, Russia}

\author{\fbox{B.I.~Khazin}}
\affiliation{Budker Institute of Nuclear Physics, SB RAS, Novosibirsk, 630090, Russia}

\author{A.N.~Kirpotin}
\affiliation{Budker Institute of Nuclear Physics, SB RAS, Novosibirsk, 630090, Russia}

\author{I.A.~Koop}
\affiliation{Budker Institute of Nuclear Physics, SB RAS, Novosibirsk, 630090, Russia}
\affiliation{Novosibirsk State University, Novosibirsk, 630090, Russia}

\author{A.A.~Korobov}
\affiliation{Budker Institute of Nuclear Physics, SB RAS, Novosibirsk, 630090, Russia}
\affiliation{Novosibirsk State University, Novosibirsk, 630090, Russia}

\author{A.N.~Kozyrev}
\affiliation{Budker Institute of Nuclear Physics, SB RAS, Novosibirsk, 630090, Russia}
\affiliation{Novosibirsk State University, Novosibirsk, 630090, Russia}
\affiliation{Novosibirsk State Technical University, Novosibirsk, 630092, Russia}

\author{E.A.~Kozyrev}
\affiliation{Budker Institute of Nuclear Physics, SB RAS, Novosibirsk, 630090, Russia}
\affiliation{Novosibirsk State University, Novosibirsk, 630090, Russia}

\author{P.P.~Krokovny}
\affiliation{Budker Institute of Nuclear Physics, SB RAS, Novosibirsk, 630090, Russia}
\affiliation{Novosibirsk State University, Novosibirsk, 630090, Russia}

\author{A.E.~Kuzmenko}
\affiliation{Budker Institute of Nuclear Physics, SB RAS, Novosibirsk, 630090, Russia}

\author{A.S.~Kuzmin}
\affiliation{Budker Institute of Nuclear Physics, SB RAS, Novosibirsk, 630090, Russia}
\affiliation{Novosibirsk State University, Novosibirsk, 630090, Russia}

\author{I.B.~Logashenko}
\affiliation{Budker Institute of Nuclear Physics, SB RAS, Novosibirsk, 630090, Russia}
\affiliation{Novosibirsk State University, Novosibirsk, 630090, Russia}

\author{P.A.~Lukin}
\affiliation{Budker Institute of Nuclear Physics, SB RAS, Novosibirsk, 630090, Russia}
\affiliation{Novosibirsk State University, Novosibirsk, 630090, Russia}

\author{A.P.~Lysenko}
\affiliation{Budker Institute of Nuclear Physics, SB RAS, Novosibirsk, 630090, Russia}

\author{K.Yu.~Mikhailov}
\affiliation{Budker Institute of Nuclear Physics, SB RAS, Novosibirsk, 630090, Russia}
\affiliation{Novosibirsk State University, Novosibirsk, 630090, Russia}

\author{I.V.~Obraztsov}
\affiliation{Budker Institute of Nuclear Physics, SB RAS, Novosibirsk, 630090, Russia}
\affiliation{Novosibirsk State University, Novosibirsk, 630090, Russia}

\author{\fbox{V.S.~Okhapkin}}
\affiliation{Budker Institute of Nuclear Physics, SB RAS, Novosibirsk, 630090, Russia}

\author{A.V.~Otboev}
\affiliation{Budker Institute of Nuclear Physics, SB RAS, Novosibirsk, 630090, Russia}

\author{E.A.~Perevedentsev}
\affiliation{Budker Institute of Nuclear Physics, SB RAS, Novosibirsk, 630090, Russia}
\affiliation{Novosibirsk State University, Novosibirsk, 630090, Russia}

\author{Yu.N.~Pestov}
\affiliation{Budker Institute of Nuclear Physics, SB RAS, Novosibirsk, 630090, Russia}

\author{A.S.~Popov}
\affiliation{Budker Institute of Nuclear Physics, SB RAS, Novosibirsk, 630090, Russia}
\affiliation{Novosibirsk State University, Novosibirsk, 630090, Russia}

\author{\fbox{G.P.~Razuvaev}}
\affiliation{Budker Institute of Nuclear Physics, SB RAS, Novosibirsk, 630090, Russia}
\affiliation{Novosibirsk State University, Novosibirsk, 630090, Russia}

\author{Yu.A.~Rogovsky}
\affiliation{Budker Institute of Nuclear Physics, SB RAS, Novosibirsk, 630090, Russia}
\affiliation{Novosibirsk State University, Novosibirsk, 630090, Russia}

\author{A.A.~Ruban}
\affiliation{Budker Institute of Nuclear Physics, SB RAS, Novosibirsk, 630090, Russia}

\author{\fbox{N.M.~Ryskulov}}
\affiliation{Budker Institute of Nuclear Physics, SB RAS, Novosibirsk, 630090, Russia}

\author{A.E.~Ryzhenenkov}
\affiliation{Budker Institute of Nuclear Physics, SB RAS, Novosibirsk, 630090, Russia}
\affiliation{Novosibirsk State University, Novosibirsk, 630090, Russia}

\author{A.V.~Semenov}
\affiliation{Budker Institute of Nuclear Physics, SB RAS, Novosibirsk, 630090, Russia}
\affiliation{Novosibirsk State University, Novosibirsk, 630090, Russia}

\author{A.I.~Senchenko}
\affiliation{Budker Institute of Nuclear Physics, SB RAS, Novosibirsk, 630090, Russia}

\author{P.Yu.~Shatunov}
\affiliation{Budker Institute of Nuclear Physics, SB RAS, Novosibirsk, 630090, Russia}

\author{Yu.M.~Shatunov}
\affiliation{Budker Institute of Nuclear Physics, SB RAS, Novosibirsk, 630090, Russia}

\author{V.E.~Shebalin}
\affiliation{Budker Institute of Nuclear Physics, SB RAS, Novosibirsk, 630090, Russia}
\affiliation{Novosibirsk State University, Novosibirsk, 630090, Russia}

\author{D.N.~Shemyakin}
\affiliation{Budker Institute of Nuclear Physics, SB RAS, Novosibirsk, 630090, Russia}
\affiliation{Novosibirsk State University, Novosibirsk, 630090, Russia}

\author{B.A.~Shwartz}
\affiliation{Budker Institute of Nuclear Physics, SB RAS, Novosibirsk, 630090, Russia}
\affiliation{Novosibirsk State University, Novosibirsk, 630090, Russia}

\author{D.B.~Shwartz}
\affiliation{Budker Institute of Nuclear Physics, SB RAS, Novosibirsk, 630090, Russia}
\affiliation{Novosibirsk State University, Novosibirsk, 630090, Russia}

\author{A.L.~Sibidanov}
\affiliation{%
 University of Victoria, Victoria, BC V8W 3P6, Canada
}%

\author{E.P.~Solodov}
\affiliation{Budker Institute of Nuclear Physics, SB RAS, Novosibirsk, 630090, Russia}
\affiliation{Novosibirsk State University, Novosibirsk, 630090, Russia}

\author{A.A.~Talyshev}
\affiliation{Budker Institute of Nuclear Physics, SB RAS, Novosibirsk, 630090, Russia}
\affiliation{Novosibirsk State University, Novosibirsk, 630090, Russia}

\author{M.V.~Timoshenko}
\affiliation{Budker Institute of Nuclear Physics, SB RAS, Novosibirsk, 630090, Russia}

\author{V.M.~Titov}
\affiliation{Budker Institute of Nuclear Physics, SB RAS, Novosibirsk, 630090, Russia}

\author{S.S.~Tolmachev}
\affiliation{Budker Institute of Nuclear Physics, SB RAS, Novosibirsk, 630090, Russia}
\affiliation{Novosibirsk State University, Novosibirsk, 630090, Russia}

\author{A.I.~Vorobiov}
\affiliation{Budker Institute of Nuclear Physics, SB RAS, Novosibirsk, 630090, Russia}

\author{Yu.V.~Yudin}
\affiliation{Budker Institute of Nuclear Physics, SB RAS, Novosibirsk, 630090, Russia}
\affiliation{Novosibirsk State University, Novosibirsk, 630090, Russia}

\author{I.M.~Zemlyansky}
\affiliation{Budker Institute of Nuclear Physics, SB RAS, Novosibirsk, 630090, Russia}

\author{D.S.~Zhadan}
\affiliation{Budker Institute of Nuclear Physics, SB RAS, Novosibirsk, 630090, Russia}

\author{Yu.M.~Zharinov}
\affiliation{Budker Institute of Nuclear Physics, SB RAS, Novosibirsk, 630090, Russia}

\author{A.S.~Zubakin}
\affiliation{Budker Institute of Nuclear Physics, SB RAS, Novosibirsk, 630090, Russia}

\collaboration{CMD-3 Collaboration}

\date{\today}

\begin{abstract}
 The cross section of the process $e^+e^-\to\pi^+\pi^-$ has been measured in the center-of-mass energy range from 0.32 to 1.2 GeV
 with the CMD-3 detector at the electron-positron collider VEPP-2000.
 The measurement is based on an integrated luminosity of about 88~pb$^{-1}$, of which 62~pb$^{-1}$ represent a complete dataset collected by CMD-3 at center-of-mass energies below 1 GeV.
 In the dominant region near the $\rho$ resonance
 a systematic uncertainty of 0.7\% was achieved.
 The implications of the presented results for the evaluation of the hadronic contribution to the anomalous magnetic moment of the muon are discussed.
\end{abstract}

\maketitle


The $e^+e^-\to\pi^+\pi^-$ process is the dominant channel of hadron production in $e^+e^-$ annihilation at center-of-mass energies, $\sqrt{s}$, below 1 GeV. The best known and most important application of the $e^+e^-\to\pi^+\pi^-$ cross section is its use for the calculation of the hadronic contribution to the anomalous magnetic moment of the muon $a_\mu=(g_\mu-2)/2$.

In the Standard Model (SM), all known interactions contribute to $a_\mu$:
\[
a_\mu^{\text{SM}} = a_\mu^{\text{QED}} + a_\mu^{\text{weak}} + a_\mu^{\text{had}},
\]
where the hadronic contribution $a_\mu^{\text{had}}$ is typically considered as the sum of the lowest order contribution, $a_\mu^{\text{had;LO}}$, also known as the hadronic vacuum polarization, and the higher order contributions. There is a difference of about 5 standard deviations between the recent experimental value of $a_\mu$~\cite{Muong-2:2023cdq} and the SM prediction \cite{Aoyama:2020ynm,Aoyama:2012wk,Aoyama:2019ryr,Czarnecki:2002nt,Gnendiger:2013pva,Davier:2017zfy,Keshavarzi:2018mgv,Colangelo:2018mtw,Hoferichter:2019mqg,Davier:2019can,Keshavarzi:2019abf,Kurz:2014wya,Melnikov:2003xd,Masjuan:2017tvw,Colangelo:2017fiz,Hoferichter:2018kwz,Gerardin:2019vio,Bijnens:2019ghy,Colangelo:2019uex,Blum:2019ugy,Colangelo:2014qya}, which has triggered a broad discussion about possible contributions from interactions beyond the SM. 

The primary method to obtain $a_\mu^{\text{had;LO}}$ employs the dispersion integral over the cross section of hadron production in $e^+e^-$ annihilation. The estimate for $a_\mu^{\text{had;LO}}$ in \cite{Aoyama:2020ynm} results from the combination of the comprehensive data-driven evaluations~\cite{Keshavarzi:2019abf,Davier:2019can,Colangelo:2018mtw}. Out of all possible hadronic channels, the $\pi^+\pi^-$ production is responsible for about 73\% of the $a_\mu^{\text{had;LO}}$ value and provides the dominant contribution to the uncertainty of the total SM prediction for $a_\mu$. The evaluations are based on the existing subpercent precision measurements of the $e^+e^-\to\pi^+\pi^-$ cross section performed on $e^+e^-$ colliders using energy scan \cite{CMD-2:2003gqi,CMD-2:2005mvb,Aulchenko:2006dxz,CMD-2:2006gxt,Achasov:2006vp,SND:2020nwa} or using the initial-state radiation (ISR) technique \cite{KLOE:2008fmq,KLOE:2010qei,KLOE:2012anl,KLOE-2:2017fda, BaBar:2012bdw,BESIII:2015equ}.
There are discrepancies between the measurements at a level of a few percent, beyond the stated uncertainties, which were accounted for by an inflation of the estimated uncertainty of $a_\mu^{\text{had;LO}}$.

Lattice QCD allows one to get an {\it ab initio} estimate of the hadronic contribution. The first sub-percent evaluation, performed by the BMW collaboration~\cite{Borsanyi:2020mff} and supported by subsequent calculations~\cite{Colangelo:2022vok,Ce:2022kxy,ExtendedTwistedMass:2022jpw,Blum:2023qou,ExtendedTwistedMassCollaborationETMC:2022sta}, led to a SM prediction $a_\mu^{\text{SM}}$ that was much closer to the experimental value, within 1.7 standard deviations.

The discrepancies in the $e^+e^-\to\pi^+\pi^-$ data and the disagreement between the data-driven and the lattice evaluations cloak the value of $a_\mu^{\text{had;LO}}$ and correspondingly $a_\mu^{\text{SM}}$ and make it impossible to search for the beyond the SM contribution to $a_\mu$ at the level allowed by the Fermilab experiment~\cite{Muong-2:2023cdq}. 

Here we present the new measurement of the $e^+e^-\to\pi^+\pi^-$ cross section $\sigma_{\pi\pi}$ performed with the CMD-3 detector at the VEPP-2000 collider. In the remainder of this Letter we will discuss the cross section in terms of the pion form factor $|F_\pi|^2$,
\begin{equation}
    \sigma_{\pi\pi}(s)=
\frac{\pi\alpha^2}{3s}\left(1-\frac{4m_\pi^2}{s}\right)^{3/2}\times|F_\pi|^2(s).
\end{equation}
A comprehensive description of data analysis and detailed discussion of results of this work are available in a companion paper~\cite{CMD-3:2023alj}.

VEPP-2000~\cite{Shatunov:2016bdv,Shwartz:2016fhe} is the symmetric electron-positron collider started operation at Budker Institute of Nuclear Physics (Novosibirsk,
Russia) in 2010. The machine covers the c.m.\ energy range from 
$\sqrt{s}=0.32$ GeV to 2.0 GeV. The unique ``round beam'' optics allows one to reach luminosities of up to $3\cdot 10^{31}$ cm$^{-2}$s$^{-1}$ at $\sqrt{s}=1$ GeV and $9\cdot 10^{31}$ cm$^{-2}$s$^{-1}$ at $\sqrt{s}=2$ GeV, which corresponds to the world's highest luminosities for the single bunch mode at this energy range. The MeV-range Compton photons produced by backscattering of the laser light on the electron beam are used for continuous monitoring of the average energy and the energy spread of the colliding beams with a systematic uncertainty of 40 keV~\cite{Abakumova:2012pn,Abakumova:2013fsa}.

The primary goal of the experiments at VEPP-2000 is to study the processes of electron-positron annihilation to hadrons, $e^+e^-\to\mathrm{hadrons}$. The detectors CMD-3\cite{Khazin:2008zz} and SND\cite{Achasov:2009zza} are installed in two interaction points of VEPP-2000. Two experiments collect data concurrently.

\begin{figure}[t]
\includegraphics[width=0.8\columnwidth]{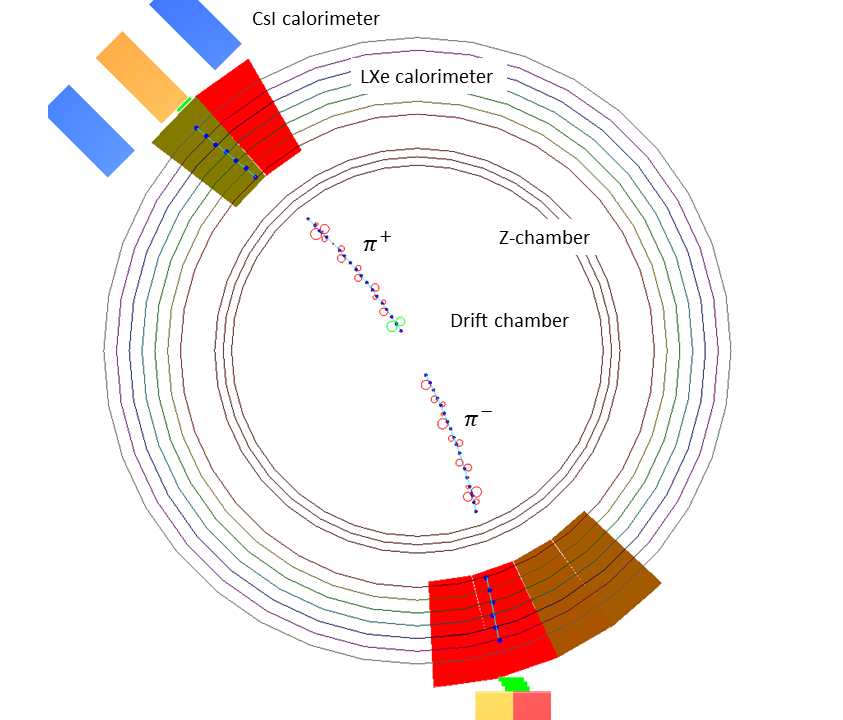}
\caption{Event display image of $e^+e^-\to\pi^+\pi^-$ event in the CMD-3 detector.}
\label{fig:event}
\hspace*{.05\textwidth}
\end{figure}

An example of the signal event $e^+e^-\to\pi^+\pi^-$ in the CMD-3 detector is shown in Fig.\ \ref{fig:event}.
The tracks of charged particles are detected by a cylindrical drift chamber with 1280 hexagonal cells with a resolution of $\approx$100 $\mu$m in the transverse plane. The coordinate along the wires, $z$, is measured with a resolution of a few mm using the charge division technique.
The $Z$ chamber is a multiwire proportional chamber with strip cathode readout, placed just outside the drift chamber, and is used for precision calibration of the $z$ measurement from the drift chamber. The tracking systems are placed inside thin superconducting solenoid (0.13X$_0$, 13 kGs).
The barrel electromagnetic calorimeter, placed outside the solenoid, consists of two systems: the inner ionization Liquid Xenon (LXe) calorimeter (about 5.4X$_0$) and the outer CsI crystal calorimeter (about 8.1X$_0$) with a time-of-flight system with sub-ns resolution located in between.
The LXe calorimeter has seven layers and uses a dual readout: the anode signals are used for a total energy deposition measurement, while the cathode strip signals provide information on a shower profile and are used for a mm-accuracy coordinate measurement. The end-cap BGO crystal calorimeter (about 13.4X$_0$) operates in the main magnetic field.
The detector is surrounded by the muon counters.

The measurement presented here is based on data taken in three distinct runs: 2013, 2018 and 2020, in a total of 209 energy points. The detector and collider conditions varied significantly between these runs, making the comparison of results between runs a valuable cross-check.


The basic idea of the measurement is straightforward. Events with two back-to-back charged pions scattered at the large angle, where the detector efficiency is the highest, are selected. The key selection criteria include the requirements for the momenta, the vertex position, the average scattering angle, the acollinearity angles $\Delta\varphi$ and $\Delta\Theta$. 

The selected sample consists of $e^+e^-\to\pi^+\pi^-$ events accompanied by $e^+e^-\to e^+e^-$ and $e^+e^-\to\mu^+\mu^-$ events
and single cosmic muons, misreconstructed as a pair of back-to-back particles originated near interaction point. The number of $e^+e^-$ pairs is used for normalization:
\begin{equation}
\label{formdef}
\left| F_{\pi }\right| ^{2}=
\left(\frac{N_{\pi\pi}}{N_{ee}} - \Delta^\text{bg}\right)
\times \frac{\sigma^{0}_{ee} (1+\delta_{ee})\varepsilon_{ee}}
{\sigma^{0}_{\pi\pi} (1+\delta_{\pi\pi})\varepsilon_{\pi\pi}},
\end{equation}
while the number of $\mu^+\mu^-$ pairs is used to check the measurement by comparing it with the ratio predicted by QED:
\begin{equation}
\label{eqnmm}
\frac{N_{\mu\mu}}{N_{ee}} = 
\frac{\sigma^{0}_{\mu\mu}(1+\delta_{\mu\mu})\varepsilon_{\mu\mu}}
{\sigma^{0}_{ee} (1+\delta_{ee})\varepsilon_{ee}}.
\end{equation}

$N_{XX}$, $X=e,\mu,\pi$, denotes here the number of $e^+e^-\to X^+X^-$ events found in the selected sample; $\sigma^{0}_{XX}$ is the lowest order cross section of the corresponding pair production in the selected solid angle range ($\sigma^{0}_{\pi\pi}$ is calculated for the pointlike pions); $\delta_{XX}$ accounts for the radiative corrections to the production cross section; $\varepsilon_{XX}$ is the detection efficiency; $\Delta^\text{bg}$ accounts for the additional background that is not directly identified in the analysis. The latter term starts to be non-negligible only at $\sqrt{s}>0.95$ GeV, since at lower energies there is practically no other background besides cosmic events and $e^+e^-\to 3\pi$ events in the narrow energy range near the $\omega(782)$ meson. Next, we will discuss the key elements of the data analysis that determine the precision of the measurement.

\paragraph{Counting number of $e^+e^-$, $\mu^+\mu^-$ and $\pi^+\pi^-$ pairs.} 
Three independent procedures were developed to measure $N_{\pi\pi}$, $N_{ee}$, and $N_{\mu\mu}$ (or combinations of these numbers). Two of them are based on the analysis of 2D distributions: the momentum of two particles ($p^+$ vs $p^-$) for the momentum-based analysis and the energy deposition in the LXe calorimeter of two particles ($E^+$ vs $E^-$) for the energy deposition-based analysis. The examples of the distributions are shown in Figs.\ 3 and 4 in \cite{CMD-3:2023alj}. The number of events of each type is extracted from the fit of the 2D distribution to a sum of shapes, predicted for each type of event.
The key feature that determines the shape of the 2D momentum distribution is the radiation of the initial and final particles. Therefore, for the momentum-based method the shapes are taken from the theoretical model [Monte Carlo (MC) generator] for $e^+e^-\to X^+X^-(\gamma)$ and then convolved with the detector response functions. In contrast, the energy deposition is largely determined by detector effects. Therefore, the shapes for the energy deposition-based method are purely empirical and are chosen to describe the data.  

The evolution of the systematic uncertainties with the beam energy is very different for the two methods. The momentum-based procedure, which is applied in our analysis at $\sqrt{s}\leq 0.9$ GeV, performs better at lower energies where the difference of $p_e$, $p_\mu$, and $p_\pi$ is large. In contrast, the energy deposition-based procedure, applied at $\sqrt{s}\geq 0.54$ GeV, is more stable at higher energies. The final ratio $N_{\pi\pi}/N_{ee}$ is the average of the results of the two methods, weighted according to their estimated systematics. The ratio $N_{\mu\mu}/N_{ee}$ is fixed to the QED prediction, adjusted for detector effects [Eq.~\eqref{eqnmm}], except for the momentum-based procedure at $\sqrt{s}\leq 0.7$ GeV, where this ratio is allowed to vary freely.

The main source of the background, cosmic muons, is considered as the fourth type of events with the corresponding shapes obtained from the data. The number of cosmic events $N_{\text{cosmic}}$ is determined in momentum-based analysis and, independently, by analyzing the distribution of the event time relative to the time of the beams collision. In average at the peak of $\rho$, the number of background events accounts for only about 0.1\% of the number of pion pairs.

The third method is based on fitting the 1D distribution of the average polar angle $dN/d\Theta$ of selected events to a sum of $dN_{XX}/d\Theta$ distributions predicted for each type of event by the corresponding theoretical model and adjusted for detector effects. The ratio $N_{\mu\mu}/N_{ee}$ is fixed to the QED prediction and the number of background events is fixed to the result of momentum-based procedure, leaving only $N_{\pi\pi}/N_{ee}$ as a free parameter. Since the statistical accuracy of the third approach is significantly inferior to the first two, it was not applied point by point, but rather used as an additional systematic check for the combined data in the energy range $\sqrt{s}=(0.7-0.82)$ GeV. The distribution and the fit are shown in Fig.\ 26 in \cite{CMD-3:2023alj}.

It should be emphasized that in the most important energy range, at the peak and the left tail of $\rho(770)$, all three methods were used and showed very good agreement at the 0.2\% level.

\paragraph{The precise determination of the polar angle of particles.}
The lowest order cross sections $\sigma^0_{XX}$ in Eq.~\eqref{formdef} depend significantly on the range of polar angle allowed in the selection of events. We have defined the allowable range as $\Theta_{\text{min}}<\Theta<\pi-\Theta_{\text{min}}$, where $\Theta$ is an average polar angle of two particles in the pair. To achieve the subpercent precision for the pion form factor, $\Theta_{\text{min}}$, which was varied between 1.4 and 1.0 rad in our analysis, should be known to $O(1\,\mathrm{mrad})$.

The polar angle for selected particles is determined by the drift chamber using the charge division method. However, this method itself cannot provide the required precision due to the insufficient long-term stability of the electronics, whose parameters change with time and temperature.
Two other detector subsystems ensure precise calibration of the charge division: the $Z$ chamber and the LXe calorimeter, both installed on the outer radius of the drift chamber. Both systems are segmented: the $Z$ chamber along the $z$ axis (the beam axis) and the LXe calorimeter along the $UV$ axes (rotated $\pm 45^0$ relative to the $z$ axis), so that the $z$ coordinate is calculated as a weighted average of fired strips.  

For the 2013 data both calibration systems were operational allowing for the cross-checks. It has been shown that the calibration of the drift chamber with either the $Z$ chamber or the LXe calorimeter allows a systematic accuracy of about 2 mrad for $\Theta$. For 2018 and 2020, only the LXe calorimeter was in operation and was used for the $z$ calibration.

\paragraph{The determination of the detection efficiencies}
The selection criteria are mainly based on the data provided by the drift chamber. The interaction of the selected $e$, $\mu$ and $\pi$ with the drift chamber materials is not exactly the same, which leads to difference in detection efficiencies $\varepsilon_{XX}$ in Eq.~\eqref{formdef}.To mitigate the potential systematic shift, only the events registered in the highly efficient part of the detector, $\Theta_{\text{min}}> 1$ rad, were used. 

Numerically, the largest source of inefficiency is the cut on the $z$ coordinate of the vertex. In order for a particle with $\Theta\approx 1$ rad to cross all wire layers, it has to originate within 5 cm of the center of the detector. The beam size $\sigma_z$ varied between 1.3 and 3.0 cm over the years of data taking, resulting in an inefficiency of up to 10\%. Special studies have shown that this inefficiency cancels out to 0.1\% or better in the ratio $\varepsilon_{\pi\pi}/\varepsilon_{ee}$.

The difference in $dE/dx$ leads to another difference in the detection efficiencies for $e$ and $\pi$ in response to the cut on the number of hit wires. The corresponding inefficiency was investigated and corrected using the data. It was found that it changes significantly, by few percent, at the edge of the allowed solid angle, $\Theta\approx 1$ rad. After the correction, no residual effect is observed at the edge when the $dN/d\Theta$ distribution is compared with the theoretical expectation, which confirms the correction.

Other potential sources of inefficiency were investigated using the test sample consisting of the particle pairs selected based on the calorimeter data. Several specific sources of inefficiency not represented with the test sample, such as the pion decays in flight, the nuclear interactions of pions, and the bremsstrahlung of electrons on the inner material of the detector, were investigated with MC and confirmed by the special data-based studies.

\paragraph{The evaluation of the radiative corrections.}
The results of the radiative correction (RC) calculations are used in two ways: to obtain $\sigma^{0}_{XX}\cdot (1+\delta_{XX})$ in Eq.~\eqref{formdef} and to obtain ideal (before detector response) shapes for the momentum-based analysis. Several effects are referred to as RC: (a) the emission of one or more $\gamma$ by electron and/or positron before the collision [initial state radiation (ISR)]; (b) the emission of one or more $\gamma$ by the final particles [final state radiation (FSR)]; (c) the interference between ISR and FSR; and (d) the virtual corrections [including vacuum polarization (VP)]. Two MC generators were used for the RC evaluation: MCGPJ~\cite{Arbuzov:2005pt} for $e^+e^-\to\pi^+\pi^-/\mu^+\mu^-$ and BabaYaga@NLO~\cite{Balossini:2006wc} for $e^+e^-\to e^+e^-/\mu^+\mu^-$. The estimated accuracy of the calculations are 0.2\% and 0.1\% respectively. Two codes use different approximations to describe the emission of multiple photons along the initial or final particles.

The generators were extensively compared for the process $e^+e^-\to e^+e^-$, which they both cover. It was shown that the calculated values of $(1+\delta_{ee})$ agree to better than 0.1\%, but the predicted spectra $d\sigma/dp^+dp^-$ differ, leading to a systematic shift in the results of momentum-based procedure.
It was observed that the spectrum predicted by BabaYaga@NLO agrees much better with the data than the one predicted by MCGPJ. The difference was attributed to the particular approximation used in MCGPJ -- that the photon jets are emitted exactly along the parent particle. The original version of MCGPJ~\cite{Arbuzov:2005pt} was modified by taking into account the angular distribution of the photons in the jet to improve the agreement with the data.

By convention, the effects of vacuum polarization are considered as part of the pion form factor; therefore, the corresponding terms are not accounted for in $\delta_{\pi\pi}$. When pion form factor is used to evaluate the hadronic contribution, it must be corrected to exclude the VP and include the FSR.

There is the chicken and egg problem related to RC: according to Eq.\ \eqref{formdef}, one needs  to know the radiation corrections $\delta_{\pi\pi}(s)$ to measure the cross section $\sigma_{\pi\pi}(s)$, but the evaluation of $\delta_{\pi\pi}(s)$ depends on the knowledge of $\sigma_{\pi\pi}(s)$. Therefore, an iterative procedure is used. We start from $\sigma_{\pi\pi}(s)$ measured in the previous experiments, use it to evaluate the RC and obtain the cross section, which is then used to re-evaluate the RC, and so on. With MC studies, it was shown that the procedure converges in 3--5 iterations. The ambiguities in the energy dependence of the cross section are added to the systematic uncertainty of the RC calculations.

The main sources of systematic uncertainty of the pion form factor measurement are listed in Table \ref{systfpi}. The estimated uncertainty depends on the energy. At the peak of the $\rho$ resonance, $\sqrt{s}=0.77$ GeV, the lowest value of 0.7\% is reached. The uncertainty increases toward lower energies up to 0.8\%, which is due to the increased contribution of pion decays in flight and particles separation. The value increases toward higher energies up to 1.6\% at $\sqrt{s}=1.0$ GeV, mainly due to the scaling of the contribution of the uncertainty of the ratio $N_{\mu\mu}/N_{ee}$ with the factor of $N_{\mu\mu}/N_{\pi\pi}$. For the 2013 data the fiducial volume contribution to the systematics was larger due to the limited performance of the tracker, which inflated the total systematic uncertainty to 0.9\% at $\sqrt{s}=0.77$ GeV and to 2.0\% at $\sqrt{s}=1.0$ GeV. 

\begin{table}[t]
\centering
\caption{\label{systfpi} Contributions to the
  systematic uncertainty of $|F_{\pi}|^2$ around $\sqrt{s}=0.77$ GeV for 2018 data. }
\small
\vskip 0.1 in
\begin{ruledtabular}
\begin{tabular}{lc} 
  Source & Contribution \\ \hline
Radiative corrections             & 0.3\% \\       
$e/\mu/\pi$ separation            & 0.2\% \\        
Fiducial volume                   & 0.5\% \\
Detector efficiency               & 0.1\% \\
Beam energy (by Compton)          & 0.1\% \\        
Bremsstrahlung loss               & 0.05\% \\
Pion nuclear interactions          & 0.2\%  \\       
Pion decays in flight             & 0.1\% \\ \hline
Total systematics                 & 0.7\% \\
\end{tabular}
\end{ruledtabular}

\end{table}

The analysis was confirmed by a series of systematic uncertainty studies. Some involved varying the selection cuts from their standard value; all results were consistent with the deviations expected due to differences in the data sample. Other checks were made by comparing the results of different separation methods and results based on datasets collected in different years. 

\begin{figure}
\includegraphics[width=\columnwidth]{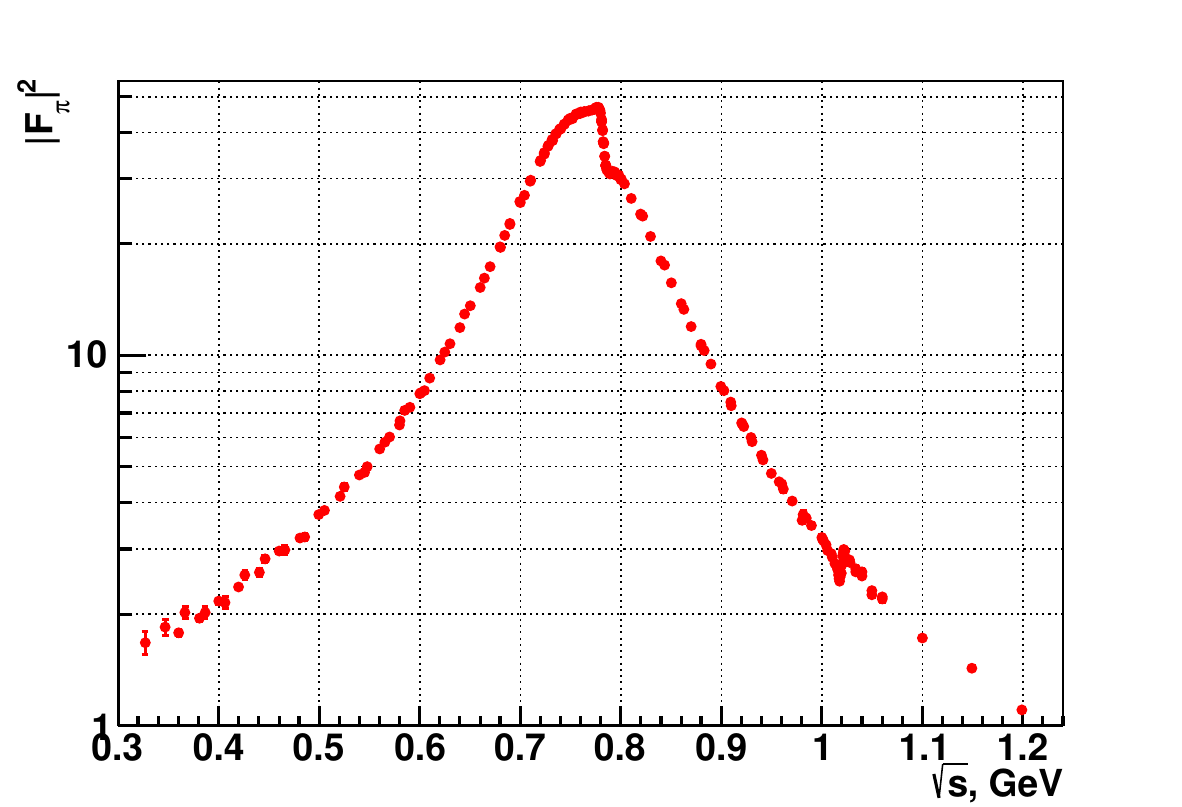}
\caption{\label{fig:result} The pion form factor measured in this work.}
\end{figure}

Two measurements performed as a byproduct of the form factor measurement provide an additional powerful consistency check. The first relates to the forward-backward charge asymmetry in $e^+e^-\to \pi^+\pi^-$~\cite{Ignatov:2022iou}. The accurate measurement of this $\sim 1$\% effect on top of the much larger asymmetry in $e^+e^-\to e^+e^-$ provides a powerful test of the accuracy of the polar angle. The energy dependence of the asymmetry observed in CMD-3 data disagreed with the theoretical prediction based on the conventional scalar QED approach~\cite{Arbuzov:2020foj}. The reason for disagreement was traced to the limitations of the scalar QED assumptions. The generalized vector-meson-dominance (GVMD) model proposed in~\cite{Ignatov:2022iou} allowed us to overcome these limitations and its prediction was found to be in agreement within the statistical uncertainties with the CMD-3 observations: the average difference between the measured and predicted asymmetry is $\delta A=(-2.9\pm 2.3)\cdot 10^{-4}$. Later these results were confirmed by an independent dispersive-based calculation~\cite{Colangelo:2022lzg}.

The second test is the measurement of $e^+e^-\to \mu^+\mu^-$ cross section, predicted by QED. It was done for momentum-based analysis for $\sqrt{s}<0.7$ GeV only, where momentum resolution of the tracking system allowed us to separate muons from other particles. The observed average ratio of the measured cross section to the QED prediction $1.0017 \pm 0.0016$ proves the consistency of the most parts of the analysis procedure, including separation procedure, detector effects, evaluation of the radiative corrections etc. 

The result of the CMD-3 pion form factor measurement is shown in Fig.\ \ref{fig:result}. 

\begin{figure}
\includegraphics[width=\columnwidth]{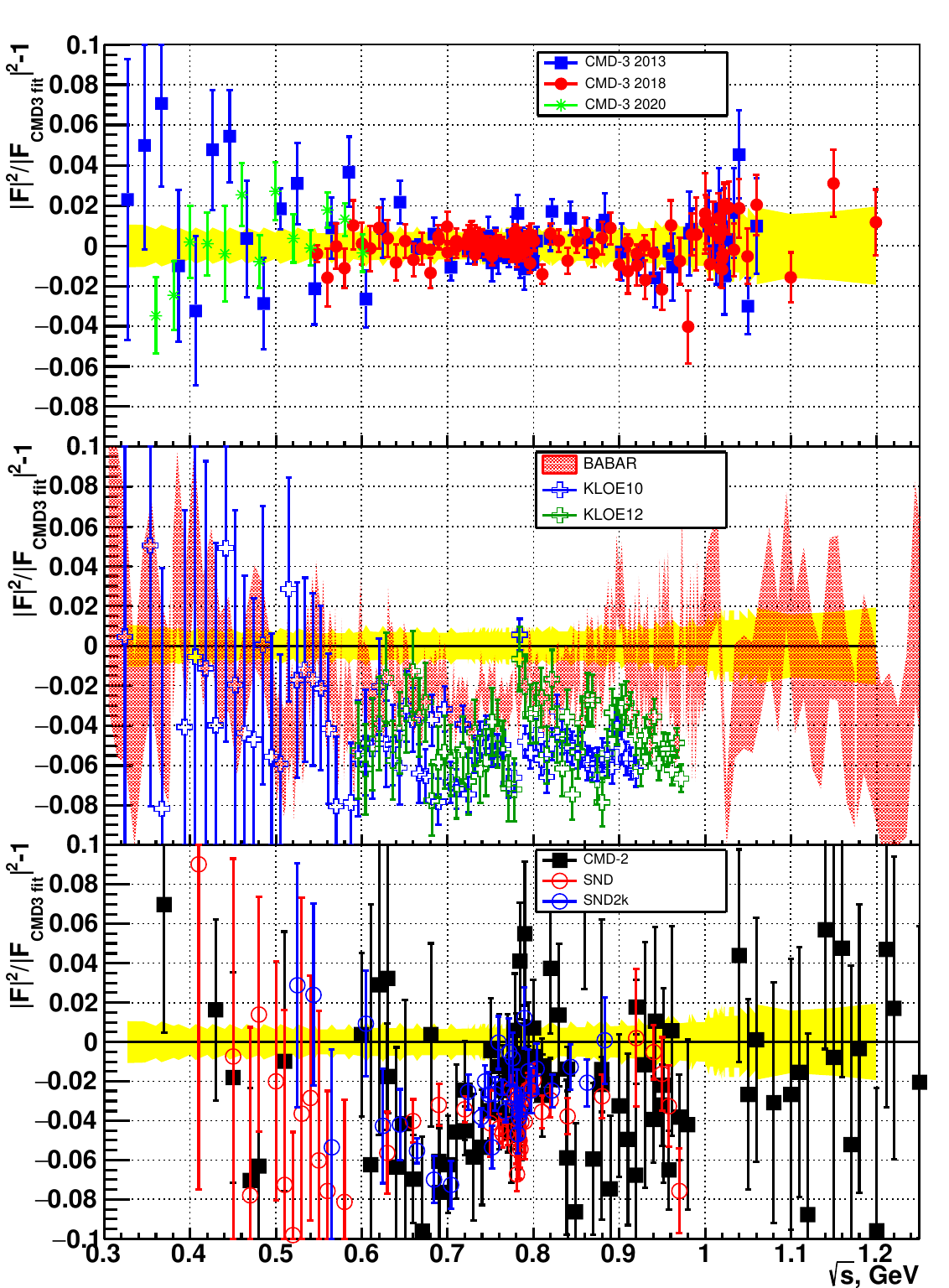}
\caption{\label{fpicomp} The relative differences between previous measurements of the pion form factor and fit of CMD-3 result, $|F_\pi|^2/|F_\pi|^2_\mathrm{CMD3\;fit}-1$. Yellow band represents CMD-3 systematic uncertainty. Top plot: CMD-3 data relative to the fit. Middle plot: ISR measurements ($BABAR$, KLOE). Bottom plot: energy scan measurements (CMD-2, SND, SND2k).}
\end{figure}

The comparison of our result to previous measurements is shown in Fig.~\ref{fpicomp}. The data points are shown relative to the fit of CMD-3 data. The band around zero reflects the systematic uncertainty of our measurement. The top plot demonstrates the distribution of our data points relative to the fit; the colors reflect three datasets discussed earlier. 
The comparison of our measurement with the most precise
ISR experiments
($BABAR$~\cite{BaBar:2012bdw}, KLOE~\cite{KLOE:2010qei,KLOE:2012anl}
) is shown in the middle plot. Two ISR measurements, BESIII~\cite{BESIII:2015equ} and 
CLEO~\cite{Xiao:2017dqv}, not shown on the plot, have somewhat larger statistical errors and consistent with both KLOE and $BABAR$. The comparison with the most precise previous energy scan experiments (CMD-2~\cite{CMD-2:2003gqi,CMD-2:2005mvb,Aulchenko:2006dxz,CMD-2:2006gxt},
SND~\cite{Achasov:2006vp} at the VEPP-2M and SND~\cite{SND:2020nwa} at the VEPP-2000, denoted as SND2k) is shown in the bottom plot. 
The new result generally shows larger pion form factor than previous experiments. The most significant difference, up to ~5\%, to other energy scan measurements is observed at the left slope of $\rho$ meson ($\sqrt{s}=0.6-0.75$ GeV).

The contribution of the $\pi^+\pi^-$ final state to the lowest order hadronic contribution $a_\mu^{\text{had;LO}}$, calculated using CMD-3 measurement, is 
\[
a_\mu^{\text{had;LO}}(2\pi; \mathrm{CMD\text{-}3}) = 5260 (42) \times 10^{-11},
\]
which should be compared to $5060 (34) \times 10^{-11}$, a value, based on the average of all previous measurements with the $\chi^2$ inflation of error to account for data inconsistencies~\cite{Aoyama:2020ynm}. Our calculation is based exclusively on CMD-3 data for $\sqrt{s}=0.327-1.2$ GeV and average of other measurements outside of this energy range. The value of the estimated error, $42\times 10^{-11}$, is completely dominated by the systematic uncertainty.

Replacing in the complete calculation of $a_\mu^{\text{had;LO}}$~\cite{Aoyama:2020ynm,Aoyama:2012wk,Aoyama:2019ryr,Czarnecki:2002nt,Gnendiger:2013pva,Davier:2017zfy,Keshavarzi:2018mgv,Colangelo:2018mtw,Hoferichter:2019mqg,Davier:2019can,Keshavarzi:2019abf,Kurz:2014wya,Melnikov:2003xd,Masjuan:2017tvw,Colangelo:2017fiz,Hoferichter:2018kwz,Gerardin:2019vio,Bijnens:2019ghy,Colangelo:2019uex,Blum:2019ugy,Colangelo:2014qya} the $\pi^+\pi^-$ contribution with our value and assuming no correlations in errors, we found the resulting Standard Model prediction for the anomalous magnetic moment of muon in a good agreement, within
0.9 standard deviations, with the most recent experimental value of $a_\mu$~\cite{Muong-2:2023cdq}:
\[
a_\mu(\textrm{exp})-a_\mu^{\text{SM}}(\mathrm{CMD\text{-}3}\;2\pi) = 49\; (55)\times 10^{-11}.
\] 

The result of this work differs significantly from the results of previous measurements, including those of the CMD-2 experiment, the predecessor of CMD-3. It should be noted that the discrepancies already observed between previous measurements, e.g., KLOE and $BABAR$, are of the same scale. The reason for these discrepancies is currently unknown and is the subject of active studies. CMD-3 and CMD-2, as well as SND, are experiments of the same type, of which CMD-3 is the next generation, featuring the improved detector performance, much more sophisticated data analysis, and a comprehensive study of systematic effects based on statistics more than an order of magnitude larger. CMD-3 and CMD-2 should be considered as independent experiments in a series of $e^+e^-\to\pi^+\pi^-$ cross-section measurements, as they share only one detector subsystem, the $Z$ chamber.

Given the recent and expected improvements in the accuracy of $a_\mu(\textrm{exp})$, the similar improvement of $a_\mu(\textrm{SM})$ is extremely important. The hadronic contribution is still a limiting factor. Some improvements are expected when the sources of the discrepancies are understood. The new measurements of the cross section of $e^+e^-\to \textrm{hadrons}$ and in particular of $e^+e^-\to \pi^+\pi^-$ with ~0.2\% systematic uncertainty are highly desirable. Such precision requires the development of next-to-next-to-leading-order MC generators for the collinear processes, which are not available at the moment.
Other ways to estimate the hadronic contribution are currently being explored, such as lattice QCD and the MUonE experiment at CERN~\cite{CarloniCalame:2015obs,Abbiendi:2016xup,Abbiendi:2677471}. All these efforts should lead to the uncertainty of $a_\mu(\textrm{SM})$ being equal to or better than $a_\mu(\textrm{exp})$.

The measured cross-section data and other byproduct results of the analysis presented in this Letter are available in the companion paper~\cite{CMD-3:2023alj}.


\bibliography{pipicmd3}

\end{document}